\documentclass[prl,aps,showpacs,twocolumn,floatfix,superscriptaddress]{revtex4}
\usepackage{amsmath,amsfonts,amssymb,graphics,graphicx,color,bm}

\begin{document}

\title{Exactly solvable pairing Hamiltonian for heavy nuclei}
\author{J. Dukelsky}
\affiliation{
  Instituto de Estructura de la Materia,
  CSIC,
  Serrano 123, E-28006 Madrid, Spain}
\author{S.~Lerma~H.}
\affiliation{
  Departamento de F\'{i}sica, Universidad Veracruzana, Xalapa, 91000, Veracruz, Mexico
  }
\author{L. M. Robledo}
\affiliation{
  Departamento de F\'{i}sica Te\'{o}rica, M\'{o}dulo 15, Universidad Aut\'{o}noma de Madrid, E-28049 Madrid, Spain
  }
\author{R. Rodriguez-Guzman}
\affiliation{
  Instituto de Estructura de la Materia,
  CSIC,
  Serrano 123, E-28006 Madrid, Spain}
\author{S. M. A. Rombouts}
\affiliation{
  Instituto de Estructura de la Materia,
  CSIC,
  Serrano 123, E-28006 Madrid, Spain}
\affiliation{Departamento de F\'{i}sica Aplicada, Universidad de Huelva, 21071 Huelva, Spain}

\begin{abstract}
We present a new exactly solvable Hamiltonian with a separable pairing
interaction and non-degenerate single-particle energies. It is derived from the hyperbolic family of Richardson-Gaudin
models and possesses two free parameters, one related to an interaction cutoff
and the other to the pairing strength. These two parameters can be adjusted
to give an excellent reproduction of  Gogny self-consistent mean-field calculations in the canonical basis.
\end{abstract}

\maketitle

Pairing is one of the most important ingredients of the effective nuclear interaction in atomic nuclei as it was early recognized by Bohr,
Mottelson, and Pines \cite{BMP} in an attempt to explain the large gaps observed in even-even nuclei. They suggested that the recently proposed Bardeen-Cooper-Schriefer (BCS) \cite{BCS} theory of superconductivity could be a useful tool in nuclear structure although care should be taken with the violation of particle number in finite nuclei. Since then, BCS or the more general Hartree-Fock-Bogoliubov (HFB) theory combined with effective or phenomenological nuclear forces has been the standard tool to describe the low energy properties of heavy nuclei. Improvements over BCS or HFB came through the restoration of broken symmetries, specially particle number projection which is still a problem not satisfactory solved with density dependent forces \cite{Luis}.
From a different perspective, Richardson found an exact solution of the constant pairing problem with non-degenerate single particle energies as early as in 1963 \cite{Rich}. Though highly schematic, the constant pairing force has been used for decades in nuclear structure with several approximations (BCS, RPA, PBCS, etc.) but scarcely resorting to the exact solution. Almost forgotten, the exact Richardson solution was recovered within the framework of ultrasmall superconducting grains \cite{Sierra1}, in which not only number projection but also pairing fluctuations were essential to describe the disappearance of superconductivity as a function of the grain size.

By combining the Richardson exact solution with the  integrable model proposed by Gaudin \cite{Gau} for quantum spin systems, it was possible to derive three families of integrable models called Richardson-Gaudin (RG) models \cite{DES}. The rational family, extensively used since then, contains the Richardson model as a particular exactly solvable Hamiltonian as well as many other exactly solvable Hamiltonians of relevance in quantum optics, cold atom physics, quantum dots, etc. \cite{DPS}. However, the other families did not find a physical realization up to very recently when it was shown that the hyperbolic family could model a $p$-wave pairing Hamiltonian in a 2 dimensional lattice \cite{Sierra}, such that it was possible to study with the exact solution an exotic phase diagram having a non-trivial topological phase and a third order quantum phase transition \cite{RDO}.
In this letter we will show that the hyperbolic family give rise to a separable pairing Hamiltonian with 2 free parameters that can be adjusted to reproduce the properties of heavy nuclei as described by a Gogny HFB treatment.

Let us start our derivation with the integrals of motion of the hyperbolic RG
model \cite{DES}, which can be  written in a compact form \cite{NuclPhysB} as
\begin{eqnarray}  \label{Rin}
R_{i}&&=S_{i}^{z}- \\
&&2\gamma \sum_{j\not=i} \left[ \frac{\sqrt{\eta _{i}\eta _{j}} }{\eta
_{i}-\eta _{j}}\left( S_{i}^{+}S_{j}^{-}+S_{i}^{-}S_{j}^{+}\right) + \frac{%
\eta _{i}+\eta _{j}}{\eta _{i}-\eta _{j}}S_{i}^{z}S_{j}^{z}\right] ,  \notag
\end{eqnarray}
where $S_{i}^{z}$, $S_{i}^{\pm}$, are the three generators of the $SU(2)_{i}$
algebra of copy $i$  with spin representation $s_{i}$ such
that $\langle S_i^2\rangle=s_i(s_i+1)$.  We assume $L$ $SU(2)$-algebra copies, $i=1,\dots,L$.
 The $L$ operators $R_{i}$ contain $L$
free parameters $\eta_{i}$ plus the strength of the quadratic term $\gamma$.
The integrals of motion (\ref{Rin}) commute among themselves and with the $z$ component of the
total spin $S^{z}=\sum_{i=1}^L S_{i}^{z}$. Therefore,
they have a common basis of eigenstates which are parametrized by the
ansatz
\begin{equation}
\left\vert \Psi_M \right\rangle =\prod\limits_{\beta =1}^{M}S_{\beta
}^{+}\left\vert \nu \right\rangle ,~S_{\beta }^{+}=\sum_{i}\frac{\sqrt{\eta
_{i}}}{\eta _{i}-E_{\beta }}S_{i}^{+} ,  \label{Psi}
\end{equation}
where $\vert \nu\rangle$ is the vacuum of the lowering operators $S_i^-\vert \nu\rangle=0$ and  the  $E_{\beta}$
($\beta=1, \cdots , M$) are the pair energies or pairons which are
determined by the condition that the  ansatz (\ref{Psi}) must satisfy the eigenvalue
equations $R_{i}\left\vert \Psi_M \right\rangle =r_{i}\left\vert \Psi_M
\right\rangle $ for every $i$.

In the pair representation of the $SU(2)$ algebra, the generators are expressed in
terms of fermion creation and annihilation operators $%
S^+_i=c^{\dagger}_i c^{\dagger}_{\overline{i}}=(S^-)^{\dagger}$, $S^z_i=
(c^{\dagger}_i c_i + c^{\dagger}_{\overline{i}} c_{\overline{i}}-1)/2$.
Each $SU(2)$ copy is associated with a single particle level $i$, with $\overline{i}$ the  time reversed partner,
and $M$ is the number of active pairs.
The vacuum $|\nu\rangle$ is defined by a
set of seniorities,  $|\nu\rangle$ = $|\nu_1,\nu_2,\ldots,\nu_l \rangle$,
where the \emph{seniority} $\nu_i=0,1$  is the number of unpaired particles in
level $i$, which determines the spin associated to the level as $s_i=(1-\nu_i)/2$.  The blocking effect of the unpaired particles reduces the number of active levels to $L_c=L-\sum_i \nu_i$.

Although any function of the integrals of motion generates an exactly
solvable Hamiltonian, we will restrict ourselves in this presentation to the
simple linear combination $H=\lambda\sum_i \eta_i R_i$.  Defining  $\lambda= \left(1+2\gamma(1-M) + \gamma L_c\right)^{-1}$, and  after  some algebraic manipulations
the Hamiltonian reduces to
\begin{equation}
H=\sum_{i} \eta_i S^z_i - G \sum_{i,i^\prime} \sqrt{\eta_i \eta_{i^\prime}} S^+_i
S^-_{i^\prime}
,  \label{Ham}
\end{equation}
where $G=2\lambda\gamma$ is a free parameter.

This Hamiltonian, expressed in a 2 dimensional momentum space basis gave rise to the celebrated $p_x+ip_y$ model of $p$-wave pairing \cite{Sierra,RDO}. However, if we interpret the parameters $\eta_i$ as single particle energies corresponding to a nuclear mean-field potential, the pairing interaction has
the unphysical behavior of increasing in strength with energy. In order to
reverse this unwanted effect we define $\eta_i = 2 (\varepsilon_i - \alpha)$, where the free parameter $\alpha$ plays the role of an energy cutoff and $%
\varepsilon_i$ is the single particle energy of the mean-field level $i$.
Making use of the pair representation of the $SU(2)$, the exactly solvable
pairing Hamiltonian (\ref{Ham}) takes the form
\begin{eqnarray}
H&=&
\sum_{i} \varepsilon _{i} \left( c_{i}^{\dagger }c_{i}+c_{\overline{i}%
}^{\dagger} c_{\overline{i}}\right)
\label{Hint} \\
& & -2 G\sum_{ii^{\prime }}\sqrt{\left( \alpha -\varepsilon _{i}\right)
\left( \alpha -\varepsilon _{i^{\prime }}\right) }c_{i}^{\dagger }c_{%
\overline{i}}^{\dagger }c_{\overline{i}^{\prime }}c_{i^{\prime }},  \notag
\end{eqnarray}

with eigenvectors given by (\ref{Psi}) and eigenvalues
\begin{equation}
E=2\alpha M+\sum_i \varepsilon_i \nu_i
+ \sum_{\beta} E_{\beta}.
\end{equation}
Here the pairons $E_{\beta}$ correspond to a solution of the set of non-linear
Richardson equations

\begin{equation}
\sum_{i}\frac{s_{i}}{\eta _{i}-E_{\beta }}-\sum_{\beta ^{\prime
}(\not=\beta )}\frac{1}{E_{\beta ^{\prime }}-E_{\beta }}=\frac{Q}{%
E_{\beta }},  \label{eq:sp_eigeneq}
\end{equation}
where $Q=\frac{1}{2 G}-\frac{L_c}{2}+M-1$. Each particular solution of Eq. (\ref{eq:sp_eigeneq}) defines a unique eigenstate (\ref{Psi}).

\begin{figure}[tH]
\begin{center}
\includegraphics[width=0.4\textwidth]{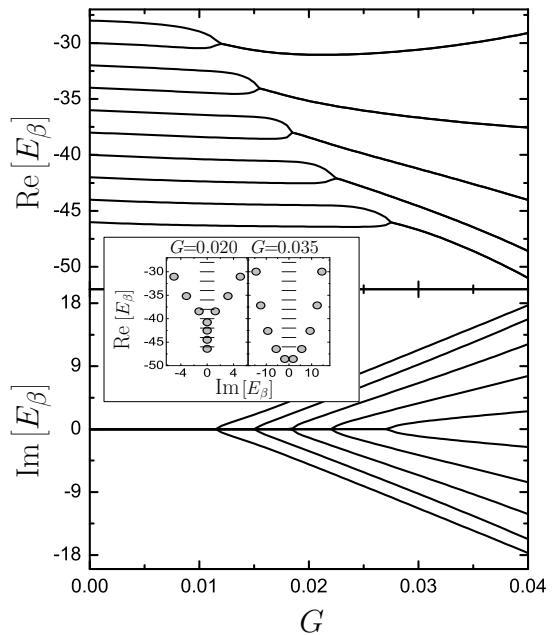}
\caption{ Real and imaginary parts of the ground-state pairons as a function of pairing strength, for a set of 24 equally spaced single-particle levels ($\varepsilon_i=i$), a cutoff $\alpha=24$ and $M=10$ pairs. The inset shows the pairon distribution in the complex plane for two different pairing strengths.
}
\label{tal}
\end{center}
\end{figure}

In order to get an insight into the solutions of (\ref{eq:sp_eigeneq}) we show in figure \ref{tal} the ground-state pairon  dependence on the pairing strength $G$ for a schematic system of $M=10$ pairs moving in a set of $L=24$ equally spaced single-particle levels ($\varepsilon_i=i$) and a  cutoff $\alpha=24$. For $G \rightarrow 0$ the pairons are all real and stay close to a set of $M$ parameters $\eta_i$ (the $M$ lowest $\eta's$ for the G.S. configuration) in order to cancel the divergence in the {\it r.h.s.} of (\ref{eq:sp_eigeneq}).
As $G$ increases the pairons move down in energy till they reach a critical value of  $G\approx 0.012$ for which the two pairons closest to the Fermi level collapse to $\eta=-30$. Immediately after they acquire an imaginary part and expand in the complex plane as a complex-conjugate pair. The same phenomenon happens to the other pairons as $G$ is further increased forming an arc in the complex plane as can be seen in the inset of Fig. \ref{tal}. Even though the behavior of the pairons resembles that of the rational model \cite{DPS}, there are qualitative differences associated to the non-constant form of the pairing interaction that will turn out to be essential for the description of heavy nuclei.

In what follows we will derive the two free parameters $G$ and $\alpha$ of the integrable Hamiltonian (\ref{Hint}) by fitting its BCS wavefuntion to a Gogny HFB calculation in the canonical basis.
The HFB calculations with the Gogny force have been carried out with the standard D1S parametrization \cite{Berger}, and the canonical basis obtained by diagonalizing the Hartree-Fock (HF) field.The pairing tensor is not exactly diagonal, but we have checked that the off diagonal contributions are much smaller than the diagonal ones. In this approximation HFB in the canonical basis is equivalent to BCS.

Due to the separable character of the integrable pairing interaction the state dependent gaps and the pairing tensor in the BCS approximation  are

\begin{figure}[ht]
\begin{center}
\includegraphics[width=0.4\textwidth]{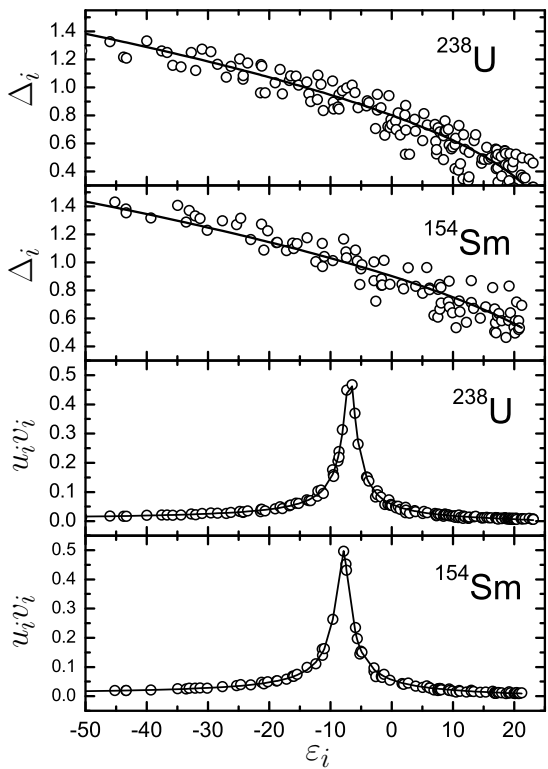}

\caption{State dependent gaps $\Delta_i$, and pairing tensor $u_i v_i$
 for protons in $^{238}$U and  $^{154}$Sm. Open circles are Gogny
HFB calculations in the canonical basis while the continuous lines are the
BCS results of the integrable Hamiltonian.}
\end{center}
\end{figure}

\begin{equation}
\Delta_i= 2 G \sqrt{ \alpha -\varepsilon_{i}} \sum_{i^{\prime}} \sqrt{%
\alpha -\varepsilon_{i^{\prime}}} u_{i^{\prime}}v_{i^{\prime}} = \Delta
\sqrt{ \alpha -\varepsilon_{i}},
\label{deli}
\end{equation}

\begin{equation}
u_i v_i= \frac{\Delta \sqrt{\alpha-\varepsilon_{i}}} {2\sqrt{(\varepsilon_{i} -\mu)^2+ (\alpha -\varepsilon_{i}) \Delta^2}}.
\label{teni}
\end{equation}

Note that the gaps $\Delta_i$ and the pairing tensor $u_i v_i$ depend on a single gap parameter $\Delta$ and have a
square root dependence on the single particle energy. Hence, the model
has a highly restricted form for both magnitudes that we will test against the Gogny
gaps $\Delta^G_i= \sum_{i^{\prime}} V_{i\overline{i},\overline{i^{\prime}%
}i^{\prime}}u^G_{i^{\prime}}v^G_{i^{\prime}}$ and pairing tensor $u^G_i v^G_i$, where $V_{i\overline{i},\overline{i^{\prime}%
}i^{\prime}}$ are the matrix elements of the Gogny force in the canonical basis and $(u^G v^G)$ is the HFB eigenvector. We take the single particle
energies $\varepsilon_i$ of the integrable Hamiltonian from the HF energies of the Gogny HFB calculations and we set up an energy cutoff of $30$
$MeV$ on top of the Fermi energy. Occupation probabilities above this cutoff are lower than $10^{-3}$ and oscillate randomly. In order to fit the two parameters of the
model $\alpha$ and $G$ and to fulfill the BCS equations for the chemical
potential $\mu$ and the gap $\Delta$, we solve the following
three coupled equations for the chemical potential $\mu$, the gap $\Delta$ and the
parameter $\alpha$:

\begin{table}[h]
\begin{center}
\begin{tabular}
[c]{|c|c|c|c|c|c|c|c|}\hline
& M & L & G & $\Delta$ & E$_{Corr}^{G}$ & E$_{Corr}^{BCS}$ & E$_{Corr}^{Exact}%
$\\\hline
$^{154}$Sm & 31 & 91 & 2.24$\times10^{-3}$ & 0.1577 & 1.3254 & 1.0164 & 2.9247
\\\hline
$^{238}$U & 46 & 148 & 1.99$\times10^{-3}$ & 0.1594 & 0.8613 & 0.5031 & 2.6511
\\\hline
\end{tabular}
\caption{\label{tab} Parameter values and correlation energies for protons in $^{154}$Sm and $^{238}$U}
\end{center}
\vspace{-0.6cm}
\end{table}

\begin{equation}
2M-L+\sum_{i}\frac{\xi _{i}}{E_{i}}=0 ,
\label{number}
\end{equation}

\begin{equation}
\sum_{i=i_{F}-n}^{i_{F}+n+1}\left[ u_{i}^{G}v_{i}^{G}-\frac{\Delta }{2}\frac{%
t_{i}}{E_{i}}\right] \frac{t_{i}\xi _{i}^{2}}{E_{i}^{3}}=0 ,
\label{uv}
\end{equation}

\begin{equation}
\sum_{i}\frac{\left( \Delta^G _{i}-\Delta \sqrt{\alpha -\varepsilon _{i}}%
\right) }{\sqrt{\alpha -\varepsilon _{i}}}=0 ,
\label{delta}
\end{equation}
where $t_i=\sqrt{\alpha-\varepsilon_i}$, $\xi_i=(\varepsilon_i - \mu)$, and the quasiparticle energy $E_i=\sqrt{{\xi_i}^2 + {\Delta_i}^2}$. Eq. (\ref{number}) is the BCS number equation that fixes the chemical potential $\mu$. Eq. (\ref{uv}) is a fitting of the Gogny pairing tensor $u^G_i v^G_i$ with respect to the gap parameters $\Delta$, {\it i.e.} we minimize  $
\sum_{i=i_{F}-n}^{i_{F}+n+1}\left( u^G_{i}v^G_{i}- u_i v_i \right) ^{2}$ with respect to $\Delta$. Here we select $n$ levels above and below the  Fermi energy in order to enhance the quality of the fit for the most correlated levels. We typically choose $n \sim 10$. Finally, Eq. (\ref{delta}) fixes the interaction cutoff $\alpha$ by minimizing the differences $\sum_{i}\left( \Delta^G _{i}-\Delta \sqrt{\alpha -\varepsilon _{i}}\right) ^{2} $ between the state dependent Gogny gaps $\Delta^G_i$ and $\Delta_i$,  with respect to $\alpha$. Once $\mu$, $\alpha$ and $\Delta$ are fixed, the pairing strength is determined from Eq. (\ref{deli},\ref{teni})
\begin{equation*}
\frac{1}{G}=\sum_{i}\frac{\left( \alpha -\varepsilon _{i}\right) }{\sqrt{\xi
_{i}^{2}+\left( \alpha -\varepsilon _{i}\right) \Delta ^{2}}}.
\end{equation*}

As a first step in ascertain the quality of the hyperbolic Hamiltonian (\ref{Hint}) to
reproduce the superfluid features of heavy nuclei, we show, in Fig. 2, the state dependent gaps $\Delta_i$ and the pairing tensor $u_i v_i$ for protons corresponding to two heavy nuclei,  $^{154}$Sm and $^{238}$U. Following the fitting procedure we consider all levels below 30$MeV$ above the Fermi energy and solve selfconsistently equations (\ref{number}-\ref{delta}) for the chemical potential $\mu$, the gap parameter $\Delta$ and the interaction cutoff $\alpha$. Fig. 2 shows a remarkable agreement between the Gogny force and the hyperbolic Hamiltonian for the pairing tensor. The Gogny state dependent gaps exhibit large fluctuations due to the details of the two-body Gogny force. However, the general trend of the gaps is very well described by the square root $\sqrt{(\alpha -\varepsilon_i)}$ of the hyperbolic model. Although $^{238}$U has 50\% more proton pairs than $^{154}$Sm the quality of the mapping is excellent for both nuclei. It is interesting to note that  the rational model, leading to the constant pairing exactly solvable Richardson Hamiltonian, has a constant gap (a horizontal line) failing completely to describe the Gogny gaps. Table I shows the number of pairs $M$, the number of active levels $L$ within the energy cutoff, the pairing strength $G$, the gap parameter $\Delta$ and the correlations energies for both nuclei.

\begin{figure}[t]
\begin{center}
\includegraphics[width=0.4\textwidth]{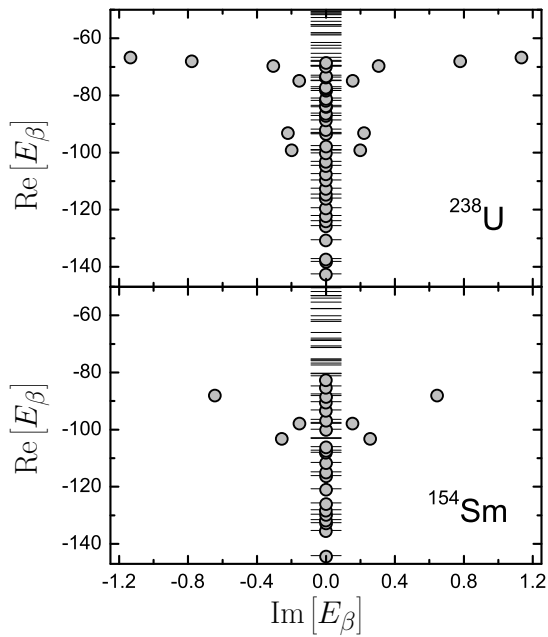}
\end{center}
\caption{Pair energies (grey circles) of the exact ground state solution for protons in $^{238}$U and $^{154}$Sm. The horizontal segments in the real axis represent the parameters $\eta_i=2(\varepsilon_i-\alpha)$.  }
\end{figure}

Once we have set up the procedure to define the parameters of the hyperbolic Hamiltonian in the BCS approximation, we are ready to explore the exact solution. For a general pairing Hamiltonian the dimension of the Hilbert space is given by the Binomial $B(L,M)$. Using the information of Table I, the dimensions are 1.98 $\times$ 10$^{24}$ for $^{154}$Sm and 4.83 $\times$ 10$^{38}$ for $^{238}$U, well beyond the limits of a large scale diagonalization. However, the integrability of the hyperbolic Hamiltonian allows us to obtain the exact solution by solving the set of $M$ non-linear coupled Richardson equations (\ref{eq:sp_eigeneq}). The exact correlation energy shown in Table I are in both nuclei considerable greater than the mean-field results, reflecting the importance of beyond mean-field quantum correlations and number fluctuations. The exact ground state wavefunction is completely determined by the position of the $M$ pairons in the complex plane. Fig. 3 shows the exact ground states for both nuclei.  Considering the structure of the pair wavefunctions (\ref{Psi}) we may argue that $^{238}$U has 4 correlated Cooper pairs, while  $^{154}$Sm has only 2. Further analysis of the Cooper pair wavefunction from the exact solutions as was carried out in \cite{Cross} for cold atoms and in \cite{Nuc} for nuclei within the rational model is straightforward but beyond the scope of this letter.

In summary, we have presented a new exactly solvable Hamiltonian with separable pairing interaction and non-degenerate single particle energies (\ref{Hint}), which arises as a particular linear combination of the hyperbolic integrals of motion (\ref{Rin}). The separable form of the pairing matrix elements could be derived from a novel Thomas-Fermi approximation for a contact interaction in a square well potential \cite{Schuck}. We have shown that the separable Hamiltonian (\ref{Hint}) with 2 free parameters is able to reproduce qualitatively the general trend of the state dependent gaps as described by the Gogny force in the canonical basis. At the same time, it reproduces accurately the HFB wavefunction represented by the pairing tensor. As such, our exactly solvable Hamiltonian is an excellent benchmark for testing approximations beyond HFB in realistic situations for even and odd nuclei. Moreover, a self-consistent HF plus exact pairing approach could be set up along the lines of Ref. \cite{Alhassid} for well bound nuclei. The inclusion of exact T=1 proton-neutron pairing within this self-consistent approach is also possible \cite{T}.

We acknowledge support from a Marie Curie Action of the European
Community Project No. 220335, the Spanish Ministry for Science and
Innovation Project No. FIS2009-07277, and FPA2009-08958,  the Mexican Secretariat
of Public Education Project PROMEP 103.5/09/4482.

\end{document}